\documentclass{tlp}

\usepackage{amsmath, amssymb}
\usepackage{epsfig}

\newtheorem{theorem}{Theorem}[section]
\newtheorem{lemma}{Lemma}[section]
\newtheorem{proposition}{Proposition}[section]

\newtheorem{example}{Example}[section]
\newtheorem{definition}{Definition}[section]

\long\def\REMOVE#1\ENDREMOVE{\message{(Commented text...)}\par}

\newcommand{\sembr}[1]{\left[\kern-0.15em\left[ #1 \right]\kern-0.15em\right]}

\title[Stabilization of Cooperative Information Agents]
{Stabilization of Cooperative Information Agents in Unpredictable
Environment: A Logic Programming Approach}

\author[P. M. Dung, D. D. Hanh, and P. M. Thang]
         {PHAN MINH DUNG, DO DUC HANH, PHAN MINH THANG\\
            Computer Science and Information Management Department\\
            Asian Institute of Technology\\
            Email: \{dung,hanh,thangphm\}@cs.ait.ac.th }

\begin{document}

\maketitle

\begin{abstract}

An information agent is viewed as a deductive database consisting
of 3 parts:
\begin{itemize}
\item an observation database containing the facts the agent has
observed or sensed from its surrounding environment.
\item an input database containing the information the agent
has obtained from other agents
\item an intensional database which is a set of rules for
computing derived information from the information stored in the
observation and input databases.
\end{itemize}

Stabilization of a system of information agents represents a
capability of the agents to eventually get correct information
about their surrounding  despite unpredictable environment changes
and the incapability of many agents to sense such changes causing
them to have temporary incorrect information. We argue that the
stabilization of a system of cooperative information agents could
be understood as the convergence of the behavior of the whole
system toward the behavior of a ``superagent'', who has the
sensing and computing capabilities of all agents combined. We show
that unfortunately, stabilization is not guaranteed in general,
even if the agents are fully cooperative and do not hide any
information from each other. We give sufficient conditions for
stabilization. We discuss the consequences of our results.

\end{abstract}


\section{Introduction}

To operate effectively in a dynamic and unpredictable environment,
agents need correct information about the environment. Often only
part of this environment could be sensed by the agent herself. As
the agent may need information about other part of the environment
that she could not sense, she needs to cooperate with other agents
to get such information. There are many such systems of
cooperative information agents operating in the Internet today. A
prominent example of such system is the system of routers that
cooperate to deliver messages from one place to another in the
Internet. One of the key characteristics of these systems is their
resilience in the face of unpredictable changes in their
environment and the incapability of many agents to sense such
changes causing them to have temporary incorrect information. This
is possible because agents in such systems cooperate by exchanging
tentative partial results to eventually converge on correct and
consistent global view of the environment. Together they
constitute a stabilizing system that allows the individual agents
to eventually get a correct view of their surrounding.

Agent communications could be classified into push-based
communications and pull-based communications. In the push-based
communication, agents periodically send information to specific
recipients. Push-based communications are used widely in routing
system, network protocols, emails, videoconferencing calls, etc. A
key goal of these systems is to guarantee that the agents have a
correct view of their surrounding. On the other hand, in the
pull-based communication, agents have to send a request for
information to other agents and wait for a reply. Until now
pull-based communications are the dominant mode of communication
in research in multiagent systems, e.g. \cite{Shoham}, \cite{Sat},
\cite{Ciam}, \cite{Kowa}, \cite{Wool97}, \cite{Wool95}. In this
paper, we consider multiagent systems where agent communications
are based on push--technologies. A prominent example of a
push-based multiagent system is the internet routing system.\\

This paper studies the problem of stabilization of systems of
cooperative information agents where an information agent is
viewed as a deductive database which consists of 3 parts:
\begin{itemize}
\item an observation database containing the facts the agent has
observed or sensed from its surrounding environment.
\item an input database containing the information the agent
was told by other agents
\item an intensional database which is a set of rules for
computing derived information from the information stored in the
observation and input databases.
\end{itemize}

It turns out that  in general, it is not possible to ensure that
the agents will eventually have the correct  information about the
environment even if they honestly exchange information and do not
hide any information that is needed by others and every change in
the environment is immediately sensed by some of the agents. We
also introduce sufficient conditions for stabilization.

The stabilization of distributed protocols has been studied
extensively in the literature
(\cite{Dijk},\cite{Flat},\cite{Schn}) where agents are defined
operationally as automata. Dijkstra \citeyear{Dijk} defined a
system as stabilizing if it is guaranteed to reach a legitimate
state after a finite number of steps regardless of the initial
state. The definition of what constitutes a legitimate state is
left to individual algorithms. Thanks to the introduction of an
explicit notion of environment, we could characterize a legitimate
state as a state in which the agents have correct information
about their environment. In this sense, we could say that our
agents are a new form of situated agents (\cite{Ros},
\cite{Brooks91}, \cite{Brooks86}) that may sometimes act on wrong
information but nonetheless will be eventually situated after
getting correct information about their surrounding. Further in
our approach, agents are defined as logic programs, and hence it
is possible for us to get general results about what kind of
algorithms could be implemented in stabilizing multiagent systems
in many applications. To the best of our knowledge, we believe
that our work is the first work on stabilization of multiagent
systems.

The rest of this paper is organized as follows. Basic notations
and definitions used in this paper are briefly introduced in
section \ref{sec:Preliminaries}. We give an illustrating example
and formalize the problem in section \ref{sec:Formalization}.
Related works and conclusions are given in section
\ref{sec:Conclusions}. Proofs of theorems are given in Appendices.

\section{Preliminaries: Logic Programs and Stable Models}\label{sec:Preliminaries}

In this section we briefly introduce the basic notations and
definitions that are needed in this paper.

We assume the existence of a Herbrand base $HB$.

A logic program is a set of ground clauses of the form:
\[
H\leftarrow L_1,\dots,L_m
\]
where $H$ is an atom from $HB$, and $L_1,\dots,L_m$ are
literals (i.e., atoms or negations of an atoms) over $HB$, $m \geq
0$. $H$ is called the head, and $L_1, \dots, L_m$ the body of the
clause.

Given a set of clauses S, the set of the heads of clauses in S is
denoted by {\it head(S)}.

Note that clauses with variables are considered as a shorthand for
the set of all their ground instantiations. Often the variables
appearing in a non-ground clause have types that are clear from
the context. In such cases these variables are instantiated by
ground terms of corresponding types.

For each atom $a$, the \textit{definition of a} is the set of all
clauses whose head is $a$.

A logic program is \textit{bounded} if the definition of every
atom is finite.

Let $P$ be an arbitrary logic program. For any set $S\subseteq
HB$, let $P^S$ be a program obtained from $P$ by deleting

\begin{enumerate}
\item each rule that has a negative literal $\neg B$ in its body
with $B\in S$, and
\item all negative literals in the bodies of the remaining rules
\end{enumerate}

$S$ is a \textit{stable model} (\cite{Gelf}) of $P$ if $S$ is the least model of $P^S$. \\

The \textit{atom dependency graph} of a logic program $P$ is a
graph, whose nodes are atoms in $HB$ and there is an edge from $a$
to $b$ in the graph iff there is a clause in $P$ whose head is $a$
and whose body contains $b$ or $\neg b$. Note that in the
literature \cite{Apt}, the direction of the link is from the atom
in the body to the head of a clause.  We reverse the direction of
the link for the ease of definition of acyclicity using the atom
dependency graph.

An atom $b$ is said to be \textit{relevant} to an atom $a$ if
there is a path from $a$ to $b$ in the atom dependency graph.

A logic program $P$ is \textit{acyclic} iff there is no infinite
path in its atom dependency graph. It is well known that
\begin{lemma}[\cite{Gelf}]
Each acyclic logic program has exactly one stable model.
\end{lemma}

\section{Examples and Problem Formalization}\label{sec:Formalization}

Routing is one of the most important problems for internetworking.
Inspired by RIP \cite{Huit}, one of the most well-known internet
routing protocols, we will develop in this section, as an example,
a multiagent system for solving the network routing problem to
motivate our work.
\begin{example}\label{ex:Routing}
Consider a network in Fig.~\ref{fig:network}. For simplicity we
assume that all links have the same cost, say 1.
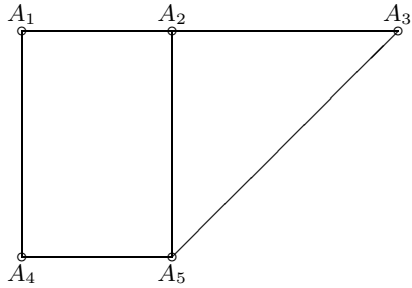
\begin{figure}[htb]
\setlength{\unitlength}{1mm}
\begin{center}
\begin{picture}(50,40)(0,0)
\linethickness{1pt} \thinlines \put(0,5){\line(0,1){30}}
\put(0,5){\line(1,0){20}} \put(0,35){\line(1,0){20}}
\put(20,35){\line(1,0){30}} \put(20,35){\line(0,-1){30}}
\put(20,5){\line(1,1){30}} \put(0,5){\circle{1}}
\put(0,35){\circle{1}} \put(20,5){\circle{1}}
\put(20,35){\circle{1}} \put(50,35){\circle{1}}
\put(0,4){\makebox(0,0)[t]{$A_4$}}
\put(20,4){\makebox(0,0)[t]{$A_5$}}
\put(0,36){\makebox(0,0)[b]{$A_1$}}
\put(20,36){\makebox(0,0)[b]{$A_2$}}
\put(50,36){\makebox(0,0)[b]{$A_3$}}
\end{picture}
\end{center}
\caption{A network example} \label{fig:network}
\end{figure}

The problem for each agent is to find the shortest paths from her
node to other nodes. The environment information an agent can
sense is the availability of links connecting to her node. The
agents use an algorithm known as ``distance vector algorithm''
(\cite{Bell}, \cite{Ford}) to find the shortest paths from their
nodes to other nodes. If the destination is directly reachable by
a link, the cost is 1. If the destination is not directly
reachable, an agent needs information from its neighbors about
their shortest paths to the destination. The agent will select the
route to the destination through a neighbor who offers a shortest
path to the destination among the agent's neighbors. Thus at any
point of time, each agent needs three kinds of information:
\begin{itemize}
\item The information about the environment, that the agent can acquire
with her sensing capability. In our example, agent $A_1$ could
sense whether the links connecting her and her neighbors $A_2,A_4$
are available.

\item The algorithm the agent needs to solve her problem. In our
example the algorithm for agent $A_1$ is represented by the
following clauses: \footnote{Contrary to the convention in Prolog,
in this paper we use lower--case letters for variables and
upper--case letters for constants.}

\[
\begin{array}{lll}
sp(A_1, A_1, 0) & \leftarrow & \\
sp(A_1, y, d)& \leftarrow & spt(A_1, y, x, d)\\
spt(A_1, y, x, d+1) & \leftarrow &  link(A_1, x), sp(x, y, d), \\
& & \quad not \,\,spl(A_1, y, d+1) \\
spl(A_1, A_1, d+1)& \leftarrow &\\
spl(A_1, y, d+1)& \leftarrow & link(A_1, x), sp(x, y, d'), d' < d
\end{array}
\]
where
\begin{description}
    \item
    $link(A_i, A_j)$ is true iff there a link from $A_i$ to
    $A_j$ in the network and the link is intact. Links are undirected,
    i.e. we identify $link(A_i, A_j)$ and $link(A_j, A_i)$.

    \item
    $sp(A_1, y, d)$ is true iff a shortest path from $A_1$ to $y$ has length $d$

    \item
    $spt(A_1, y, x, d)$ is true iff the length of shortest paths from
    $A_1$ to $y$ is $d$ and there is a shortest path from $A_1$ to $y$
    that goes through $x$ as the next node after $A_1$

    \item
    $spl(A_1, y, d)$ is true iff there is a path from $A_1$ to
    $y$ whose length is less than $d$.
\end{description}

\item
The information the agent needs from other agents. For agent $A_1$
to calculate the shortest paths from her node to say $A_3$, she
needs the information about the length of the shortest paths from
her neighbors $A_2$, and $A_4$ to $A_3$, that means she needs to
know the values $d$, $d'$ such that $sp(A_2$, $A_3, d)$, $sp(A_4,
A_3, d')$ hold.

\end{itemize}
\end{example}

\subsection{Problem Formalization}

The agents are situated in the environment. They may have
different accessibility to the environment depending on their
sensing capabilities. The environment is represented by a set of
(ground) environment atoms, whose truth values could change in an
unpredictable way.

\begin{definition}

An agent is represented by a quad-tuple
\[
A= (IDB,HBE, HIN, \delta)
\]
where

\begin{itemize}

\item
$IDB$, the intensional database, is an acyclic logic program.

\item
$HBE$ is the set of all (ground) environment atoms whose
truth values the agent could sense, i.e. $a\in HBE$ iff $A$ could
discover instantly any change in the truth value of $a$ and update
her extensional database accordingly.

\item
$HIN$ is the set of all atoms called input atoms,
whose truth values the agent must obtain from other agents.

No atom in $HIN \cup HBE$ appears in the head of the clauses in
$IDB$ and $HIN \cap HBE=\emptyset$.

\item
$\delta$ is the initial state of the agent.

\end{itemize}

\end{definition}


\begin{definition} An agent state is a pair
$ \sigma = (EDB, IN)$ where
\begin{itemize}

\item
$EDB\subseteq HBE$ represents what the agent has sensed from the
environment. That means for each $a\in HBE$, $a\in EDB$ iff $a$ is
true.

\item
$IN \subseteq HIN$, the input database of $A$, represents the set
of information $A$ has obtained from other agents, i.e. $a\in IN$
iff $A$ was told that $a$ is true.

\end{itemize}

Given a state $\sigma = (EDB, IN)$, the {\it stable model} of $A=
(IDB,HBE,HIN, \delta)$ at $\sigma$ is defined as the stable model
of $IDB \cup EDB\cup IN$. Note that $\delta$ and $\sigma$ could be
different states.
\end{definition}


\begin{example}[Continuation of the network routing example]
\label{ex:network} Imagine that initially the agents have not sent
each other any information and all links are intact. In this
situation, agent $A_1$ is represented as follows:

\begin{itemize}
\item
$IDB_1$ contains the clauses shown in Example \ref{ex:Routing}.

\item
$HBE_1= \{link(A_1, A_2), link(A_1, A_4)\}$

\item
$HIN_1$ consists of ground atoms of the form
\[
    sp(A_2,Y,D),\, sp(A_4,Y,D)
\]
\noindent where $Y\in \{A_2,\dots,A_5\}$ and $D$ is a positive
integer.

\item
The initial state $\delta_1=(EDB_{1,0}, IN_{1,0})$ where
\[
\begin{array}{l}
EDB_{1,0} = \{link(A_1, A_2), link(A_1, A_4)\}\\
IN_{1,0}= \emptyset
\end{array}
\]

\end{itemize}

\end{example}

\begin{definition}
A cooperative multiagent system is a collection of $n$ agents $
(A_1, \dots, A_n), $ with $A_i$= $(IDB_i$,$HBE_i$, $HIN_i$,
$\delta_i)$ such that the following conditions are satisfied
\begin{itemize}
\item
for each atom $a$, if $a \in head(IDB_i) \cap head(IDB_j)$
then $a$ has the same definition in $IDB_i$ and $IDB_j$.

\item
for each agent $A_i$, $HIN_i\subseteq
\bigcup\limits_{\begin{array}{l}j=1
\end{array}}^n
(head(IDB_j)\cup HBE_j)$

\item No environment atom appears in the head of clauses in the
intentional database of any agent, i.e. for all i,j: $HBE_i \cap
head(IDB_j) = \emptyset$.
\end{itemize}


For each agent $A_i$ let $HB_i= head(IDB_i) \cup HBE_i \cup
HIN_i$.
\end{definition}

\subsection{Agent Communication and Sensing}
\label{subsec:CommunicationSensing}

Let $A_i= (IDB_i,HBE_i, HIN_i, \delta_{i})$  for $1\leq i \leq n$.
We say that $A_i$ {\bf depends} on $A_j$ if $A_i$ needs input from
$A_j$, i.e. $HIN_i \cap (head(IDB_j) \cup HBE_j) \neq \emptyset$.
The {\bf dependency} of $A_i$
on $A_j$ is defined to be the set $D(i,j)=HIN_i \cap (head(IDB_j) \cup HBE_j)$. \\

\noindent As we have mentioned before, the mode of communication
for our agents corresponds to the ``push--technology''. Formally,
it means that if $A_i$ depends on $A_j$ then $A_j$ will
periodically send $A_i$ a set $S = D(i,j) \cap M_j$ where $M_j$ is
the stable model of $A_j$. When $A_i$ obtains S, she knows that
each atom $a\in D(i,j)\setminus S$ is false with respect to $M_j$.
Therefore she will update her input database $IN_i$ to
$Upa_{i,j}(IN_i,S)$ as follows
\[
Upa_{i,j}(IN_i,S) = (IN_i \setminus D(i,j)) \cup S
\]
Thus her
state has changed from $\sigma_i=(EDB_i, IN_i)$ to
$\sigma_i'=$ $(EDB_i,$ $Upa_{i,j}(IN_i,S))$ accordingly.\\

\noindent An environment change is represented by a pair $C =
(T,F)$ where $T$ (resp. $F$) contains the atoms whose truth values
have changed from false (resp. true) to true (resp. false).
Therefore, given an environment change $(T,F)$, what $A_i$ could
sense of this change, is captured by the pair $(T_i,F_i)$ where
$T_i = T \cap HBE_i$ and $F_i = F \cap HBE_i$. Hence when a change
$C = (T,F)$ occurs in the environment, agent $A_i$ will update her
sensing database $EDB_i$ to $Upe_i(EDB_i,C)$ as follows:
\[
Upe_i(EDB_i,C) = (EDB_i \setminus F_i) \cup T_i
\]

The state of agent $A_i$ has changed from $\sigma_i=(EDB_i, IN_i)$
to $\sigma_i'=$ $(Upe_i(EDB_i,C),$ $IN_i)$ accordingly.

\subsection{Semantics of Multiagent Systems}
\label{subsec:DynamicSemantics}

Let
\[
{\cal A} = (A_1,\ldots,A_n)
\]
with
\[
A_i= (IDB_i,HBE_i,HIN_i, \delta_{i})
\]
be a multiagent system. $(\delta_1,\dots,\delta_n)$ is
called the \textit{initial state} of $\mathcal{A}$.\\

A \textit{state} of $\cal A$ is defined as
\[
    \triangle=(\sigma_1,\ldots,\sigma_n)
\]
such that $\sigma_i$ is a state of agent $A_i$.\\

There are two types of transitions in a multiagent system. A
environment transition  happens when there is a change in the
environment which  is sensed by a set of agents and causes these
agents to update their extensional databases accordingly. A
communication transition   happens when an agent sends information
to another agent and causes the later to update her
input database accordingly.\\

For an environment change $C=(T, F)$, let $S_C$ be the set of
agents which could sense parts of $C$, i.e.
\[
S_C = \{ A_i \,|\, HBE_i \cap (T \cup F) \neq \emptyset\}
\]

\begin{definition}
\label{def:transition} Let $\triangle=(\sigma_1,\ldots,\sigma_n)$,
$\triangle'=(\sigma'_1,\ldots,\sigma'_n)$ be states of
$\mathcal{A}$ with $\sigma_i=(EDB_i,IN_i)$,
$\sigma'_i=(EDB'_i,IN'_i)$.
\begin{enumerate}
\item \label{trans:1}A environment transition
\[
\triangle\xrightarrow[]{C} \triangle'
\]
caused by an environment change $C = (T,F)$ is defined as follows
\begin{enumerate}
\item for every agent $A_k$ such that $ A_k \not\in S_C$: $\sigma_k = \sigma'_k$, and
\item for each agent $A_i \in S_C$:
\begin{itemize}
\item $EDB'_i = Upe_i(EDB_i,C)$,
\item $IN_i' = IN_i$.
\end{itemize}
\end{enumerate}

\item \label{trans:2}
A communication transition
\[
\triangle\xrightarrow{j\leadsto i}\triangle'
\]
caused by agent $A_j$ sending information to agent $A_i$, where
$A_i$ depends on $A_j$, is defined as follows:
\begin{enumerate}
\item For all $k$ such that $ k \neq i$: $\sigma_k = \sigma'_k$
\item $EDB_i' = EDB_i$ and $IN_i' = Upa_{i,j}(IN_i,S)$ where
$S = D(i,j) \cap M_j$ and $M_j$ is the stable model of $A_j$ at
$\sigma_j$.
\end{enumerate}
\end{enumerate}

We often simply write  $\triangle \rightarrow \triangle'$ if there
is a transition $\triangle\xrightarrow[]{C} \triangle'$ or
$\triangle\xrightarrow{j\leadsto i}\triangle'$.

\end{definition}

\begin{definition}
\label{def:run} A \textbf{run} of a multiagent system $\cal A$ is
an infinite sequence
\[
    \triangle_0 \rightarrow \triangle_1 \rightarrow \ldots \rightarrow
    \triangle_m \rightarrow \ldots
\]
such that
\begin{itemize}
\item
$\triangle_0$ is the initial state of $\mathcal{A}$
and for all agents $A_i, A_j$ such that $A_i$ depends on $A_j$ the
following condition is satisfied:
\begin{center}
\textit{For each $h$, there is a $k \geq h$ such that $\triangle_k
\xrightarrow{j \leadsto i} \triangle_{k+1}$}\\
\end{center}
The above condition is introduced to capture the idea that agents
periodically send the needed information to other agents.

\item
There is a point $h$ such that at every $k\geq h$ in the run,
there is no more environment change.\\
\end{itemize}
\end{definition}


For a run $\mathcal{R}=\triangle_0 \rightarrow \triangle_1
\rightarrow \ldots\rightarrow\triangle_k\rightarrow\ldots$ where
$\triangle_k=(\sigma_{1,k},\dots,\sigma_{n,k})$ we often refer to
the stable model of $A_i$ at state $\sigma_{i,k}$ as \textit{the
stable model of $A_i$ at point $k$} and denote it by $M_{i,k}$.

\begin{example}\label{ex:CyclicProgram}

Consider the following multiagent system
\[
\mathcal{A} = (A_1, A_2)
\]
where
\[
\begin{array}{lllllll}
IDB_1 & = & \{a \leftarrow b,c& &IDB_2 & = & \{ b \leftarrow a,d\\
& & f \leftarrow a\} & & & & \,\, \,\,b \leftarrow e\}\\
HBE_1 & = & \{c\} & & HBE_2 & =& \{d,e\}\\
HIN_1 & = & \{b\} & &  HIN_2 & = &\{a\}\\
EDB_{1,0} & = & \{c\} & & EDB_{2,0} & = & \{d,e\}\\
IN_{1,0} & = & \emptyset & & IN_{2,0} & = & \emptyset
\end{array}
\]

Consider the following run $\mathcal{R}$, where the only
environment change occurs at point $2$ such that the truth value
of $e$ becomes false:
\[
\begin{array}{l}
\triangle_0 \xrightarrow{2 \leadsto 1} \triangle_1 \xrightarrow{1
\leadsto 2} \triangle_2 \xrightarrow{(\emptyset, \{e\})}
\triangle_{3}\xrightarrow{1 \leadsto 2}
\triangle_{4}\xrightarrow{2 \leadsto 1}\triangle_{5}\dots
\end{array}
\]

The states and stable models of $A_1$ and $A_2$ at points $0$,
$1$, $2$, $3$, and $4$ are as follows \\
\begin{center}
\begin{tabular}{|c|c|c|c|c|c|c|}
  \cline{1-7}
      & \multicolumn{3}{c|}{$A_1$} & \multicolumn{3}{c|}{$A_2$} \\   \cline{1-7}
  $k$ & $EDB$ & $IN$ & Stable Model & $EDB$ & $IN$ & Stable Model \\ \cline{1-7}
  $0$ & $\{c\}$ & $\emptyset$ & $\{c\}$ & $\{d,e\}$ & $\emptyset$ & $\{b,d,e\}$ \\
  $1$ & $\{c\}$ & $\{b\}$ & $\{a,b,c,f\}$ & $\{d,e\}$ & $\emptyset$ & $\{b,d,e\}$ \\
  $2$ & $\{c\}$ & $\{b\}$ & $\{a,b,c,f\}$ & $\{d,e\}$ & $\{a\}$ & $\{a,b,d,e\}$ \\
  $3$ & $\{c\}$ & $\{b\}$ & $\{a,b,c,f\}$ & $\{d\}$ & $\{a\}$ & $\{a,b,d\}$ \\
  $4$ & $\{c\}$ & $\{b\}$ & $\{a,b,c,f\}$ & $\{d\}$ & $\{a\}$ & $\{a,b,d\}$ \\
  \cline{1-7}
\end{tabular}
\end{center}

\end{example}

\begin{example}[Continuation of example \ref{ex:network}]\label{ex:app1}

Consider the following run $\mathcal{R}$ of the multiagent system
given in Example \ref{ex:network}.
\[
\begin{array}{l}
\triangle_0 \xrightarrow{2\leadsto 1} \triangle_1
\xrightarrow{(\emptyset,\{link(A_1,A_2)\})}\triangle_2\rightarrow
\dots
\end{array}
\]

Initially, all links are intact and all inputs of agents are
empty, i.e. $IN_{i,0}=\emptyset$ for $i=1,\dots,5$. At point $0$
in the run, agent $A_2$ sends to agent $A_1$ information about
shortest paths from her to other agents. At point $1$ in the run,
the link between $A_1$ and $A_2$ is down.

The information (output) an agent needs to send to other agents
consists of shortest paths from her to other agents. Thus from the
stable model of an agent we are interested only in this output.

Let $SP_{i,k}$ be the set $\{sp(A_i, Y, D)|sp(A_i, Y, D)\in
M_{i,k}\}$ where $M_{i,k}$ is the stable model of $A_i$ at point
$k$. $SP_{i,k}$ denotes the output of $A_i$ at point $k$. It is
easy to see that if there is a transition
$\triangle_k\xrightarrow{j\leadsto i}\triangle_{k+1}$, then $A_j$
sends to $A_i$:
\[
S=D(i,j)\cap M_{j,k} = SP_{j,k}
\]

At point $0$, $A_1$ and $A_2$ have the following states and
outputs:
\[
\begin{array}{lll}
EDB_{1,0} & = & \{link(A_1,A_2), link(A_1,A_4)\} \\
IN_{1,0} & = & \emptyset\\
SP_{1,0} & = &\{sp(A_1, A_1,0)\}\\
EDB_{2,0} & = & \{link(A_2, A_1), link(A_2, A_3),link(A_2, A_5)\}\\
IN_{2,0} & = & \emptyset \\
SP_{2,0} & = & \{sp(A_2, A_2,0)\}
\end{array}
\]

$A_2$ sends $S$ to $A_1$ in the transition $\triangle_0
\xrightarrow{2\leadsto 1} \triangle_1$ where
\[
S= SP_{2,0} = \{sp(A_2, A_2, 0)\}
\]
Thus
\[
IN_{1,1}=Upa_{1,2}(IN_{1,0},S)=Upa_{1,2}(\emptyset,S)=S= \{sp(A_2,
A_2, 0)\}
\]

The environment change $C=(\emptyset,\{link(A_1, A_2)\})$ at point
$1$ is sensed by $A_1$ and $A_2$. The states of $A_1$ and $A_2$
are changed as follows:
\[
\begin{array}{lll}
  IN_{1,2} & = & IN_{1,1} \\
  EDB_{1,2} & = & Upe_1(EDB_{1,1},C) = (EDB_{1,1}\setminus \{link(A_1, A_2)\})\cup\emptyset\\
            & = & \{link(A_1,A_4)\}\\
  IN_{2,2} & = & IN_{2,1}\\
  EDB_{2,2}  & = & Upe_2(EDB_{2,1},C) = (EDB_{2,1}\setminus \{link(A_1, A_2)\})\cup\emptyset\\
  & = & \{link(A_2,A_3), link(A_2, A_5)\}\\
\end{array}
\]

The following tables show the states and outputs of $A_1$ and
$A_2$ at points $0$, $1$, and $2$ respectively.\\

\begin{center}
\begin{tabular}{|c|c|c|c|}
  \cline{1-4}
  \multicolumn{4}{|c|}{$A_1$}\\ \cline{1-4}
  $k$& $EDB$ & $IN$ & $SP$ \\ \cline{1-4}
  $0$ & $\{link(A_1,A_2), link(A_1,A_4)\}$ & $\emptyset$ & $\{sp(A_1,A_1,0)\}$ \\
  $1$ & $\{link(A_1,A_2), link(A_1,A_4)\}$ & $\{sp(A_2,A_2,0)\}$ & $\{sp(A_1,A_1,0), sp(A_1,A_2,1)\}$ \\
  $2$ & $\{ link(A_1,A_4)\}$ & $\{sp(A_2,A_2,0)\}$ & $\{sp(A_1,A_1,0)\}$   \\ \cline{1-4}
\end{tabular}
\begin{tabular}{|c|c|c|c|}
  \multicolumn{4}{c}{}\\
  \cline{1-4}
  \multicolumn{4}{|c|}{$A_2$}\\ \cline{1-4}
  $k$& $EDB$ & $IN$ & $SP$
  \\\cline{1-4}
  $0$ & $\{link(A_2,A_1), link(A_2,A_3), link(A_2,A_5)\}$ & $\emptyset$ & $\{sp(A_2,A_2,0)\}$ \\
  $1$ & $\{link(A_2,A_1), link(A_2,A_3), link(A_2,A_5)\}$ & $\emptyset$ & $\{sp(A_2,A_2,0)\}$ \\
  $2$ & $\{link(A_2,A_3), link(A_2,A_5)\}$ & $\emptyset$ & $\{sp(A_2,A_2,0)\}$ \\ \cline{1-4}
\end{tabular}
\end{center}

\end{example}

\subsection{Stabilization}

Consider a  superagent whose sensing capability and problem
solving capability are the combination of the sensing capabilities
and problem solving capabilities of all agents, i.e. this agent
can sense any change in the environment  and her intensional
database is the union of the intensional databases of all other
agents. Formally, the superagent of a multiagent system
\[
\mathcal{A}=(A_1,\dots, A_n)
\]
where
\[
A_i=(IDB_i, HBE_i,HIN_i,\delta_i), ~\delta_i=(EDB_i,IN_i)
\]
is represented by
\[
P_\mathcal{A}=(IDB_\mathcal{A}, \delta)
\]
where
\begin{itemize}
\item $IDB_\mathcal{A} = IDB_{1} \cup \dots\cup IDB_{n}$
\item $\delta$, the initial state of $P_\mathcal{A}$, is equal to $EDB_1 \cup \dots \cup
EDB_n$
\end{itemize}

The superagent actually represents the multiagent system in the
ideal case where each agent has obtained the correct information
for its input atoms.

\begin{example}[Continuation of Example \ref{ex:CyclicProgram}]
\label{ex:CyclicProgram1} Consider the multiagent system in
Example \ref{ex:CyclicProgram}. At point $0$, the superagent
$P_\mathcal{A}$ is represented as follows:
\begin{itemize}
\item
$IDB_\mathcal{A}$ consists of the following clauses:
\[
\begin{array}{l}
a\leftarrow b,c \qquad f\leftarrow a \qquad b\leftarrow a,d \qquad
b\leftarrow e
\end{array}
\]

\item
$\delta=\{c,d,e\}$.
\end{itemize}

\end{example}

\begin{example}[Continuation of Example \ref{ex:app1}]
\label{ex:SuperAgent}

Consider the multiagent system in Example \ref{ex:app1}.
Initially, when all links between nodes are intact, the superagent
$P_{\cal A}$ is represented as follows:
\begin{itemize}
\item
$IDB_\mathcal{A}$ consists of the following clauses:
\[
\begin{array}{lll}
sp(x, x, 0) & \leftarrow & \\
sp(x, y, d) & \leftarrow & spt(x, y, z, d)\\
spt(x, y, z, d+1) & \leftarrow &  link(x, z), sp(z, y, d), \\
& & \quad not \,\,spl(x, y, d+1)\\
spl(x, x, d+1) & \leftarrow & \\
spl(x, y, d+1) & \leftarrow & link(x, z), sp(z, y, d'), d' < d
\end{array}
\]

\item The initial state
\[
\begin{array}{ll}
\delta= \{ & link(A_1,A_2), link(A_1, A_4), link(A_2, A_3),\\
          & link(A_2, A_5), link(A_3, A_5), link(A_4, A_5)\}
\end{array}
\]
\end{itemize}

Note that the possible values of variables $x$, $y$, $z$ are
$A_1$, $A_2$, $A_3$, $A_4$, $A_5$.
\end{example}

\begin{definition} Let $\mathcal{A}$ be a multiagent system.

The \textbf{I/O graph} of $\mathcal{A}$ denoted by $G_\mathcal{A}$
is a graph obtained from the atom dependency graph of its
superagent's intensional database $IDB_\mathcal{A}$ by removing
all nodes that are not relevant for any input atom in
$HIN_1\cup\dots\cup HIN_n$.

$\mathcal{A}$ is \textbf{IO-acyclic} if there is no infinite path
in its I/O graph $G_\mathcal{A}$.

$\mathcal{A}$ is \textbf{bounded} if $IDB_\mathcal{A}$ is bounded.

$\mathcal{A}$ is \textbf{IO-finite} if its I/O graph is finite.
\end{definition}


\begin{example}

The atom dependency graph of $IDB_\mathcal{A}$ and the I/O-graph
$G_\mathcal{A}$ of the multiagent system in Examples
\ref{ex:CyclicProgram} and \ref{ex:CyclicProgram1} is given in
Fig. \ref{fig:I/O-graph}.

\begin{figure}[htb]
\centering \epsfig{file=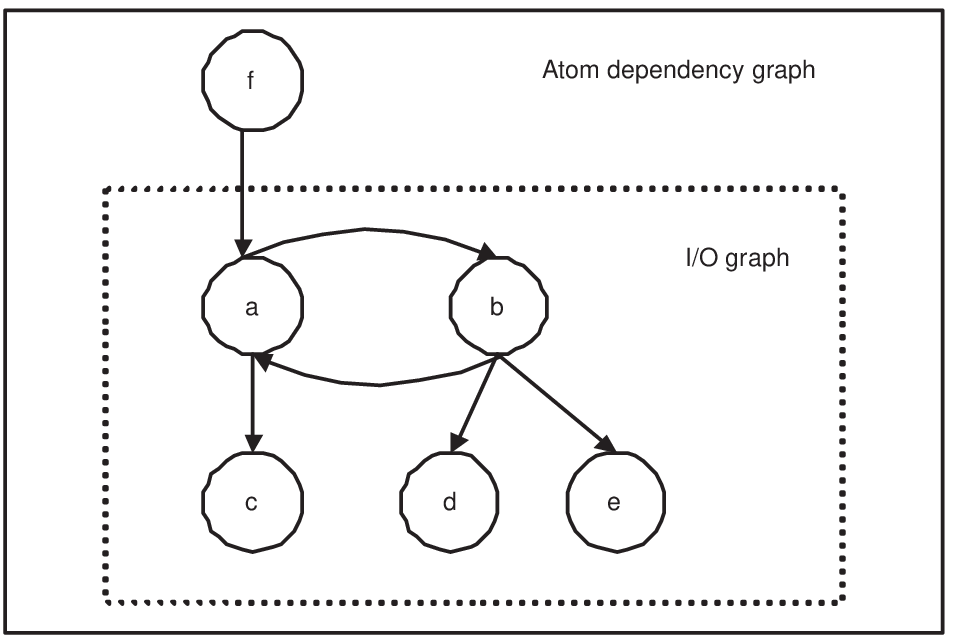, height=7cm} \caption{The
atom dependency graph and I/O graph} \label{fig:I/O-graph}
\end{figure}

It is obvious that the multiagent system in Examples
\ref{ex:CyclicProgram} and \ref{ex:CyclicProgram1} is bounded but
not IO-acyclic and the multiagent system in Examples
\ref{ex:Routing}, \ref{ex:network}, \ref{ex:app1} and
\ref{ex:SuperAgent} is IO-acyclic and bounded.

\end{example}

\begin{proposition}
\label{prop:acyclic} If a multiagent system $\mathcal{A}$ is
\textbf{IO-acyclic} then $IDB_\mathcal{A}$ is acyclic.
\end{proposition}

\begin{proof}

Suppose $IDB_\mathcal{A}$ is not acyclic. There is an infinite
path $\eta$ in its atom dependency graph starting from some atom
$a$. There is some agent $A_i$ such that $a\in HB_i$. Since
$IDB_i$ is acyclic, every path in its atom dependency graph is
finite. $\eta$ must go through some atom $b\in IN_i$ to outside of
$A_i$'s atom dependency graph. Clearly starting from $b$, all
atoms in $\eta$ are relevant to $b$. The infinite path of $\eta$
starting from $b$ is a path in the I/O graph $G_\mathcal{A}$.
Hence $G_\mathcal{A}$ is not acyclic. Contradiction!
\end{proof}

\begin{definition} Let $\mathcal{R}=\triangle_0\rightarrow\dots\triangle_k\rightarrow\dots$
be a run and $M_{i,k}$ be the stable model of $A_i$ at point $k$.
\begin{enumerate}
\item $\mathcal{R}$ is \textbf{convergent} for an atom $a$ if either
of the following conditions is satisfied.
\begin{itemize}
  \item There is a point $h$ such that at every point $k\geq h$,
  for every agent $A_i$ with $a\in HB_i= head(IDB_i) \cup HBE_i\cup HIN_i$,
    \[
    a\in M_{i,k}
    \]
  In this case we write $Conv(\mathcal{R},a)=true$
  \item There is a point $h$ such that at every point $k\geq h$,
  for every agent $A_i$ with $a\in HB_i$,
        \[
        a\not\in M_{i,k}
        \]
    In this case we write $Conv(\mathcal{R},a)=false$
\end{itemize}

\item $\mathcal{R}$ is \textbf{convergent} if it is convergent
for each atom.

\item $\mathcal{R}$ is {\textbf{strongly convergent}} if it is
convergent and there is a point $h$ such that at every point
$k\geq h$, for every agent $A_i$, $M_{i,k}=M_{i,h}$.
\end{enumerate}
\end{definition}


It is easy to see that strong convergence implies convergence.
Define
\[ Conv(\mathcal{R}) = \{a\,|\, Conv(\mathcal{R},a)=true\}
\]
as the \textbf{convergence model} of $\mathcal{R}$.

Let $\mathcal{R}=\triangle_0 \rightarrow \triangle_1 \rightarrow
\ldots\rightarrow\triangle_k\rightarrow\ldots$ be a run where
$\triangle_k=(\sigma_{1,k},\dots,\sigma_{n,k})$ with
$\sigma_{i,k}=(EDB_{i,k}, IN_{i,k})$. As there is a point $h$ such
that the environment does not change after $h$, it is clear that
$\forall k\geq h: EDB_{i,k}=EDB_{i,h}$. The set $EDB=\bigcup
\limits_{i=1}^n EDB_{i,h}$ is called \textbf{the stabilized
environment} of $\mathcal{R}$.

\begin{definition}
\begin{itemize}
\item A multiagent system is said to be {\bf weakly stabilizing} if
every run $\mathcal{R}$ is convergent, and its convergence model
$Conv(\mathcal{R})$ is a stable model of $P_\mathcal{A}$ in the
stabilized environment of $\mathcal{R}$, i.e. $Conv(\mathcal{R})$
is a stable model of $IDB_\mathcal{A}\cup EDB$ where $EDB$ is the
stabilized environment of $\mathcal{R}$.
\item A multiagent system is said to be {\bf stabilizing} if it is
weakly stabilizing and all of its runs are strongly convergent.
\end{itemize}
\end{definition}


\begin{theorem}
\label{theo:1} IO-acyclic and bounded multiagent systems are
weakly stabilizing.
\end{theorem}
\begin{proof}
See \ref{app:ProofTheorem1}.
\end{proof}

\vspace{0.5cm}

Unfortunately, the above theorem does not hold for more general
class of multiagent systems as the following example shows.

\begin{example}[Continuation of example \ref{ex:CyclicProgram} and
\ref{ex:CyclicProgram1}]
\label{ex:CyclicProgram2}

Consider the multiagent system $\mathcal{A}$ and run $\mathcal{R}$
in Example \ref{ex:CyclicProgram}. It is obvious that
$\mathcal{A}$ is bounded but not IO-acyclic.

For every point $k\geq 4$, $M_{1,k}=\{a,b,c,f\}$,
$M_{2,k}=\{a,b,d\}$. $Conv(\mathcal{R})=\{a,b,c,d,f\}$. The
stabilized environment of $\mathcal{R}$ is $EBD=\{c,d\}$. The
stable model of $P_\mathcal{A}$ in the stabilized environment of
$\mathcal{R}$ is $\{c,d\}$, which is not the same as
$Conv(\mathcal{R})$. Hence the system is not weakly stabilizing.
\end{example}

Boundedness is very important for the weak stabilization of
multiagent systems. Consider a multiagent system in the following
example which is IO-acyclic, but not bounded.

\begin{example}\label{ex:InfiniteHB}

Consider the following multiagent system
\[ \mathcal{A} = (A_1,
A_2)
\]
where
\[
\begin{array}{lll}
IDB_1 = \{ q \leftarrow \neg r(x) & & IDB_2= \{r(x+1) \leftarrow s(x)\\
\qquad \qquad \,\,s(x) \leftarrow r(x)\} & &\qquad \qquad \,\,r(0)\leftarrow\}\\
HBE_1 =  \{\} & & HBE_2 =  \{\}\\
HIN_1=  \{r(0),r(1),\dots\} & & HIN_2= \{s(0), s(1), \dots \}\\
EDB_{1,0}=\emptyset \,\,\,\, IN_{1,0}=\emptyset & &
EDB_{2,0}=\emptyset \,\,\,\, IN_{2,0}=\emptyset
\end{array}
\]

Since $HBE=HBE_1\cup HBE_2=\emptyset$, for every run $\mathcal{R}$
the stabilized environment of $\mathcal{R}$ is empty. The stable
model of $P_\mathcal{A}$ in the stabilized environment of
$\mathcal{R}$ is the set $\{r(0), r(1),\dots\}$$\cup$$\{s(0),
s(1),\dots\}$. It is easy to see that for each run, the agents
need to exchange infinitely many messages to establish all the
values of $r(x)$. Hence for every run $\mathcal{R}$, for every
point $h\geq 0$ in the run: $q \in M_{1,h}$, but $q$ is not in the
stable model of $P_\mathcal{A}$ in the stabilized environment of
$\mathcal{R}$. Thus the system is not weakly stabilizing.
\end{example}

Are the boundedness and IO-acyclicity sufficient to guarantee the
stabilization of a multiagent system? The following example shows
that they are not.

\begin{example}[Continuation of Example \ref{ex:app1} and \ref{ex:SuperAgent}]
\label{ex:StabCond}

Consider the multiagent system in Example \ref{ex:network}.
Consider the following run $\mathcal{R}$ with no environment
change after point $6$.
\begin{align}
\triangle_0\xrightarrow{5\leadsto 2} \triangle_1 \xrightarrow{5\leadsto 4} \triangle_2 \xrightarrow{2\leadsto 1}\\
\triangle_{3} \xrightarrow{(\emptyset, \{link(A_1, A_2)\})} \triangle_4\xrightarrow{4\leadsto 1}\\
\triangle_5 \xrightarrow{(\emptyset, \{link(A_4, A_5)\})}
\triangle_6 \xrightarrow{1\leadsto 4}  \label{p3}\\
\triangle_7 \xrightarrow{4\leadsto 1}\triangle_8
\rightarrow\dots\label{p4}
\end{align}

Initially all links in the network are intact. The states and
outputs of agents are as follows:
\begin{itemize}
\item $EDB_{1,0}=\{link(A_1, A_2),link(A_1, A_4)\}$,

$EDB_{2,0}=\{link(A_2, A_1),link(A_2, A_3),link(A_2,A_5)\}$

$EDB_{3,0}=\{link(A_3, A_2),link(A_3, A_5)\}$.

$EDB_{4,0}=\{link(A_4, A_1),link(A_4, A_5)\}$.

$EDB_{5,0}=\{link(A_5, A_2),link(A_5, A_3),link(A_5, A_4)\}$.
\item $IN_{i,0}=\emptyset$ for $i=1,\dots, 5$.
\item $SP_{i,0}=\{sp(A_i, A_i,0)\}$ for $i=1, \dots, 5$.
\end{itemize}

Recall that $SP_{i,k}$ denotes the output of $A_i$ at point $k$
and is defined as follows:
\[
SP_{i,k} = \{sp(A_i, Y, D)|sp(A_i, Y, D)\in M_{i,k}\}
\]

The following transitions occur in $\mathcal{R}$:
\begin{itemize}
\item At point $0$, $A_5$ sends $SP_{5,0}=\{sp(A_5,A_5,0)\}$ to
$A_2$. This causes the following changes in the input and output
of $A_2$:
\[
\begin{array}{ll}
IN_{2,1}=& \{sp(A_5, A_5,0)\}\\
SP_{2,1}=& \{sp(A_2,A_2,0), sp(A_2, A_5,1)\}\\
\end{array}
\]
\item At point $1$, $A_5$ sends $SP_{5,1}=\{sp(A_5,A_5,0)\}$ to
$A_4$. This causes the following changes in the input and output
of $A_4$:
\[
\begin{array}{ll}
IN_{4,2}=& \{sp(A_5, A_5,0)\}\\
SP_{4,2}=& \{sp(A_4,A_4,0), sp(A_4, A_5,1)\}\\
\end{array}
\]
\item At point $2$, $A_2$ sends $SP_{2,2}=\{sp(A_2, A_2,0),
sp(A_2,A_5,1)\}$ to $A_1$. This causes the following changes in
the input and output of $A_1$:
\[
\begin{array}{ll}
IN_{1,3}=& \{sp(A_2, A_2,0), sp(A_2, A_5,1)\}\\
SP_{1,3}=& \{sp(A_1,A_1,0), sp(A_1, A_2,1), sp(A_1, A_5,2)\}\\
\end{array}
\]
\item At point $3$, the link between $A_1$ and $A_2$ is down as shown in Fig.~\ref{fig:network1}.
\begin{figure}[htb]
\setlength{\unitlength}{1mm}
\begin{center}
\begin{picture}(50,40)(0,0)
\linethickness{1pt} \thinlines \put(0,5){\line(0,1){30}}
\put(0,5){\line(1,0){20}} \put(20,35){\line(1,0){30}}
\put(20,35){\line(0,-1){30}} \put(20,5){\line(1,1){30}}
\put(0,5){\circle{1}} \put(0,35){\circle{1}}
\put(20,5){\circle{1}} \put(20,35){\circle{1}}
\put(50,35){\circle{1}} \put(0,4){\makebox(0,0)[t]{$A_4$}}
\put(20,4){\makebox(0,0)[t]{$A_5$}}
\put(0,36){\makebox(0,0)[b]{$A_1$}}
\put(20,36){\makebox(0,0)[b]{$A_2$}}
\put(50,36){\makebox(0,0)[b]{$A_3$}}
\end{picture}
\end{center}

\caption{The network after the link between $A_1$ and $A_2$ is
down} \label{fig:network1}
\end{figure}
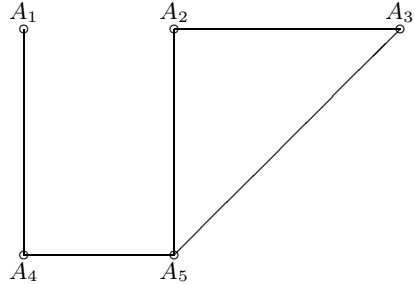
This causes the following changes in the states and outputs of
$A_1$ and $A_2$:
\[
\begin{array}{ll}
EDB_{1,4}=\{link(A_1, A_4)\} & EDB_{2,4}=\{link(A_2, A_3), link(A_2, A_5)\}\\
IN_{1,4}= \{sp(A_2, A_2,0), sp(A_2, A_5,1)\} & IN_{2,4}=\{sp(A_5,A_5,0)\}\\
SP_{1,4}= \{sp(A_1,A_1,0)\} & SP_{2,4}=\{sp(A_2,A_2,0), sp(A_2, A_5,1)\}\\
\end{array}
\]

\item At point $4$, $A_4$ sends $SP_{4,4}=\{sp(A_4, A_4,0),
sp(A_4,A_5,1)\}$ to $A_1$. This causes the following changes in
the input and output of $A_1$:
\[
\begin{array}{ll}
IN_{1,5}=& \{sp(A_2, A_2,0), sp(A_2, A_5,1), sp(A_4, A_4,0), sp(A_4, A_5,1)\}\\
SP_{1,5}=& \{sp(A_1,A_1,0), sp(A_1, A_4,1), sp(A_1, A_5,2)\}\\
\end{array}
\]

\item At point $5$, the link between $A_4$ and $A_5$ is down as shown in Fig.~\ref{fig:network2}.
\begin{figure}[htb]
\setlength{\unitlength}{1mm}
\begin{center}
\begin{picture}(50,40)(0,0)
\linethickness{1pt} \thinlines \put(0,5){\line(0,1){30}}
\put(20,35){\line(1,0){30}} \put(20,35){\line(0,-1){30}}
\put(20,5){\line(1,1){30}} \put(0,5){\circle{1}}
\put(0,35){\circle{1}} \put(20,5){\circle{1}}
\put(20,35){\circle{1}} \put(50,35){\circle{1}}
\put(0,4){\makebox(0,0)[t]{$A_4$}}
\put(20,4){\makebox(0,0)[t]{$A_5$}}
\put(0,36){\makebox(0,0)[b]{$A_1$}}
\put(20,36){\makebox(0,0)[b]{$A_2$}}
\put(50,36){\makebox(0,0)[b]{$A_3$}}
\end{picture}
\end{center}

\caption{The network after the link between $A_4$ and $A_5$ is
down} \label{fig:network2}
\end{figure}
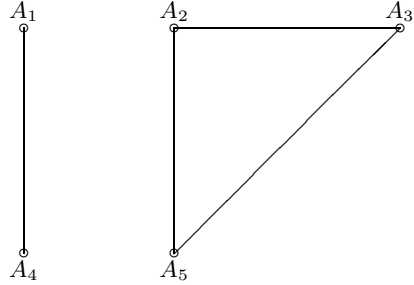
This causes the following changes in the states and outputs of
$A_4$ and $A_5$:
\[
\begin{array}{ll}
EDB_{4,6}=\{link(A_4, A_1)\} & EDB_{5,6}=\{link(A_5, A_2), link(A_5, A_3)\}\\
IN_{4,6}= \{sp(A_5, A_5, 0)\} & IN_{5,6}=\emptyset\\
SP_{4,6}= \{sp(A_4, A_4, 0)\} & SP_{5,6}=\{sp(A_5,A_5,0)\}\\
\end{array}
\]

\item At point $6$, $A_1$ sends $SP_{1,6}=\{sp(A_1, A_1,0),
sp(A_1,A_5,2)\}$ to $A_4$. This causes the following changes in
the input and output of $A_4$:
\[
\begin{array}{ll}
IN_{4,7}=& \{sp(A_5,A_5,0), sp(A_1, A_1,0), sp(A_1, A_5,2)\}\\
SP_{4,7}=& \{sp(A_4,A_4,0), sp(A_4, A_1,1), sp(A_4, A_5,3)\}\\
\end{array}
\]

Note that at point $6$, $sp(A_1, A_5,2)\in M_{1,6}$, i.e. the
length of the shortest path from $A_1$ to $A_5$ equals to $2$, is
wrong. But $A_1$ sends this information to $A_4$. Now the length
of the shortest paths to $A_5$ of agents $A_1$, and $A_4$ equal to
$2$, and $3$ respectively (i.e. $sp(A_1,A_5,2)\in M_{1,7}$ and
$sp(A_4,A_5,3)\in M_{4,7}$, are all wrong. Later on $A_1$ and
$A_4$ exchange wrong information, increase the shortest paths to
$A_5$ after each round by $2$ and go into an infinite loop.
\end{itemize}

The states and outputs of $A_1$ and $A_4$ at points $0\rightarrow
8$ are shown in Fig.~\ref{StateA1} and Fig.~\ref{StateA4}
respectively.

\begin{figure}[htb]
\begin{center}
\begin{tabular}{|c|c|c|c|}
  \cline{1-4}
  $k$ & $EDB$ & $IN$ & $SP$ \\ \cline{1-4}
  $0$ & $\{link(A_1,A_2),$ & $\emptyset$ & $\{sp(A_1, A_1,0)\}$ \\
      & $\,\,link(A_1,A_4)\}$ &  &  \\
  \cline{1-4}
  $1$ & $\{link(A_1,A_2),$ & $\emptyset$ & $\{sp(A_1, A_1,0)\}$ \\
      & $\,\,link(A_1,A_4)\}$ & & \\
  \cline{1-4}
  $2$ & $\{link(A_1,A_2),$ & $\emptyset$ & $\{sp(A_1, A_1,0)\}$ \\
      & $\,\,link(A_1,A_4)\}$ & & \\
  \cline{1-4}
  $3$ & $\{link(A_1,A_4)\}$ & $\{sp(A_2,A_2,0), sp(A_2, A_5,1)\}$& $\{sp(A_1, A_1,0),sp(A_1, A_2,1),$\\
      &                     &                 & $sp(A_1, A_5,2)\}$ \\
  \cline{1-4}
  $4$ & $\{link(A_1,A_4)\}$ & $\{sp(A_2,A_2,0), sp(A_2, A_5,1)\}$ & $\{sp(A_1, A_1,0)\}$\\
  \cline{1-4}
  $5$ & $\{link(A_1,A_4)\}$ & $\{sp(A_2,A_2,0), sp(A_2, A_5,1),$ & $\{sp(A_1, A_1,0), sp(A_1, A_4,1),$\\
      &                     & $\,\,sp(A_4, A_4,0),sp(A_4,A_5,1)\}$ & $sp(A_1, A_5, 2)\}$\\
  \cline{1-4}
  $6$ & $\{link(A_1,A_4)\}$ & $\{sp(A_2,A_2,0), sp(A_2, A_5,1),$ & $\{sp(A_1, A_1,0),sp(A_1, A_4,1),$\\
      &                     & $\,\,sp(A_4, A_4,0),sp(A_4,A_5,1)\}$ & $ sp(A_1, A_5,2)\}$\\
  \cline{1-4}
  $7$ & $\{link(A_1,A_4)\}$ & $\{sp(A_2,A_2,0), sp(A_2, A_5,1),$ & $\{sp(A_1, A_1,0),sp(A_1, A_4,1),$\\
      &                  & $\,\,sp(A_4, A_4,0), sp(A_4, A_5, 1)\}$ & $sp(A_1, A_5,2)\}$ \\
  \cline{1-4}
  $8$ & $\{link(A_1,A_4)\}$ & $\{sp(A_2,A_2,0), sp(A_2, A_5,1),$ & $\{sp(A_1, A_1,0),sp(A_1, A_4,1),$\\
      &                    & $\,\,sp(A_4, A_4,0), sp(A_4, A_5, 3)\}$ & $ sp(A_1, A_5,4)\}$\\
  \cline{1-4}
\end{tabular}
\end{center}
\caption{State and output of $A_1$} \label{StateA1}
\end{figure}

\begin{figure}[htb]
\begin{center}
\begin{tabular}{|c|c|c|c|}
  \cline{1-4}
  $k$ & $EDB$ & $IN$ & $SP$ \\ \cline{1-4}
  $0$ & $\{link(A_4,A_1),$ & $\emptyset$ & $\{sp(A_4, A_4,0)\}$\\
       & $\,\,link(A_4,A_5)\}$ &  & \\
  \cline{1-4}
  $1$ & $\{link(A_4,A_1),$ & $\emptyset$ & $\{sp(A_4, A_4,0)\}$\\
      & $\,\,link(A_4,A_5)\}$ &  & \\
  \cline{1-4}
  $2$ & $\{link(A_4,A_1),$ & $\{sp(A_5,A_5,0)\}$ & $\{sp(A_4, A_4,0),sp(A_4,A_5,1)\}$\\
      & $\,\,link(A_4,A_5)\}$ &  & \\
  \cline{1-4}
  $3$ & $\{link(A_4,A_1),$ & $\{sp(A_5,A_5,0)\}$ & $\{sp(A_4, A_4,0),sp(A_4,A_5,1)\}$\\
      & $\,\,link(A_4,A_5)\}$ &  &\\
  \cline{1-4}
  $4$ & $\{link(A_4,A_1),$ &  $\{sp(A_5,A_5,0)\}$ & $\{sp(A_4, A_4,0),sp(A_4,A_5,1)\}$\\
      & $\,\,link(A_4,A_5)\}$ &   & \\
  \cline{1-4}
  $5$ & $\{link(A_4,A_1),$ & $\{sp(A_5,A_5,0)\}$ & $\{sp(A_4, A_4,0),sp(A_4,A_5,1)\}$\\
      & $\,\,link(A_4,A_5)\}$ &  & \\
  \cline{1-4}
  $6$ & $\{link(A_4,A_1)\}$ & $\{sp(A_5,A_5,0)\}$ & $\{sp(A_4, A_4,0)\}$\\
  \cline{1-4}
  $7$ & $\{link(A_4,A_1)\}$ & $\{sp(A_5,A_5,0), sp(A_1,A_1,0),$ & $\{sp(A_4, A_4,0),sp(A_4, A_1,1),$\\
      &                     & $ sp(A_1,A_5,2)\}$ & $sp(A_4, A_5,3\}$\\
  \cline{1-4}
  $8$ & $\{link(A_4,A_1)\}$ & $\{sp(A_5,A_5,0), sp(A_1,A_1,0),$ & $\{sp(A_4, A_4,0),sp(A_4, A_1,1),$\\
      &                     & $ sp(A_1,A_5,2)\}$ & $sp(A_4, A_5,3\}$\\
  \cline{1-4}
\end{tabular}
\end{center}
\caption{State and output of $A_4$} \label{StateA4}
\end{figure}

\end{example}

This example shows that
\begin{theorem}
IO-acyclicity and boundedness are not sufficient to guarantee the
stabilization of a multiagent system.
\end{theorem}

As we have pointed out before, the routing example in this paper
models the popular routing RIP protocol that has been widely
deployed in the internet.  Example \ref{ex:StabCond} shows that
RIP  is not stabilizing. In configuration \ref{fig:network2}, the
routers at the nodes $A_1,A_4$ go into a loop and continuously
change the length of the shortest paths from them to $A_5$ from 2
to infinite. This is because the router at node $A_1$ believes
that the shortest path from it to $A_5$ goes through $A_4$ while
the router at $A_4$ believes that the shortest path from it to
$A_5$ goes through $A_1$. None of them realizes that there is no
more connection between them and $A_5$. \footnote{This is one of
the key reasons why RIP, a very simple internet routing protocol,
is gradually replaced by OSPF, a much more complex routing
protocol \cite{Huit}}. The above theorem generalizes this insight
to multiagent systems. The conclusion is that in general it is not
possible for an agent to get correct information about its
environment if this agent can not sense all the changes in the
environment by itself and has to rely on the communications with
other agents. This is true even if all the agents involved are
honest and do not hide their information.

Obviously, if a multiagent system is IO-acyclic and IO-finite,
every agent would obtain complete and correct information after
finitely many exchanges of information with other agents. The
system is stabilizing. Hence

\begin{theorem}
\label{theo:2} IO-acyclic and IO-finite multiagent systems are
stabilizing.
\end{theorem}

\begin{proof}
See \ref{app:ProofTheorem2}.
\end{proof}

\section{Related Works and Conclusions}\label{sec:Conclusions}

There are many research works on multiagent systems where agents
are formalized in terms of logic programming such as \cite{Ciam},
\cite{Kowa}, \cite{Sat}. An agent in our framework could be viewed
as an abductive logic program as in \cite{Ciam}, \cite{Sat} where
atoms in the input database could be considered as abducibles.
Satoh and Yamatomo formalized speculative computation with
multiagent belief revision. The semantics of multiagent systems,
which is defined based on belief sets and the union of logic
programs of agents, is similar to our idea of ``superagent''. An
agent in \cite{Ciam} is composed of two modules: the Abductive
Reasoning Module (ARM), and the Agent Behaviour Module (ABM).
Agents are grouped within bunches according to the requirements of
interaction between agents. The coordination (collaboration) of
agents is implicitly achieved through the semantics of the
consistency operators. In both works (\cite{Ciam} and \cite{Sat})
the communication for agents is based on pull-technologies. The
authors did not address the stabilization issue of multiagent
systems. Sadri, Toni and Torroni in \cite{Sadri} used a
logic-based framework for negotiation to tackle the resource
reallocation problem via pull-based communication technology and
the solution is considered as ``stabilization'' property.

In this paper, we consider a specific class of cooperative
information agents without considering effects of their actions on
the environment e.g. in \cite{Ciam}, \cite{Kowa}, \cite{Sat}. We
are currently working to extend the framework towards this
generalized issue.

In this paper, a logic programming based framework for cooperative
multiagent systems is introduced, and the stabilization of
multiagent systems is then formally defined. We introduced
sufficient conditions in general for multiagent systems under
which the stabilization is guaranteed. We showed that IO-acyclic
and bounded multiagent systems are weakly stabilizing. But
IO-acyclicity and boundedness are not sufficient to guarantee the
stabilization of a multiagent system. We showed that IO-acyclic
and IO-finite multiagent systems are stabilizing. Unfortunately
these conditions are strong. So it is not an easy task to ensure
that agents eventually get right information in the face of
unpredictable changes of the environment.

Our research is inspired by the network routing applications. As
the RIP (\cite{Hedr}, \cite{Huit}) is very simple and had been
widely accepted and implemented. But the RIP has many limitations
such as the bouncing effect, counting to infinity, looping, etc.
Many versions and techniques of the RIP have been introduced to
reduce undesired features of the RIP, but the problem could not be
solved thoroughly. With logic programming approach, we showed in
this paper, the main reason is that in the RIP, the computation of
the overall problem solving algorithm is distributed over the
network, while the logic program which represents the routing
algorithm is not IO-finite, the stabilization of the system is
thus not guaranteed. It is also a reason why most experts prefer
the OSPF (\cite{Moy}, \cite{Huit}), which is much more complicated
and sophisticated protocol, to the RIP for network routing.

We have assumed that information sent by an agent is obtained
immediately by the recipients. But communications in real networks
always have delay and errors in transmissions. We believe that the
results presented in this paper could also be extended for the
case of communication with delay and errors.

In this paper communications for agents are based on
push-technologies. It is interesting to see how the results could
be extended to multiagent systems whose communication is based on
pull-technologies (\cite{Sat}, \cite{Ciam}).

\appendix

\section{Proof of theorem \ref{theo:1}} \label{app:ProofTheorem1}

First it is clear that the following lemma holds.
\begin{lemma}
\label{lemma:app10} Let $M$ be a stable model of a logic program
$P$. For each atom $a$: $a\in M$ iff there is a clause
$a\leftarrow Bd$ in $P$ such that $M\models Bd$.
\end{lemma}

Given an IO-acyclic and bounded multiagent system
$\mathcal{A}=(A_1, \dots, A_n)$. By proposition
\ref{prop:acyclic}, $IDB_\mathcal{A}$ is acyclic.

Let
\[
\mathcal{R}=\triangle_0\rightarrow\dots\rightarrow\triangle_h\rightarrow\dots
\]
be a run of $\mathcal{A}$ such that after point $h$ there is no
more change in the environment. The stabilized environment of
$\mathcal{R}$ is $EDB=EDB_{1,h}\cup\dots\cup EDB_{n,h}$. Let
$\sembr{P_\mathcal{A}}$ be the stable model of $P_\mathcal{A}$ in
the stabilized environment of $\mathcal{R}$, i.e. the stable model
of $IDB_\mathcal{A}\cup EDB$.

The height of an atom $a$ in the atom dependency graph of
$P_\mathcal{A}$ denoted by $\pi(a)$ is the length of a longest
path from $a$ to other atoms in the atom dependency graph of
$P_\mathcal{A}$. Since $IDB_\mathcal{A}$ is acyclic, there is no
infinite path in the atom dependency graph of
$P_\mathcal{A}$. From the boundedness of $IDB_\mathcal{A}$, $\pi(a)$ is finite.\\

Theorem \ref{theo:1} follows directly from the following lemma.

\begin{lemma}
\label{lemma:app11}

For every atom $a$, $\mathcal{R}$ is convergent for $a$ and
$conv(\mathcal{R},a)=true$ iff $a\in \sembr{P_\mathcal{A}}$.
\end{lemma}

It is easy to see that lemma \ref{lemma:app11} follows immediately
from the following lemma.

\begin{lemma}
\label{lemma:app12} For every atom $a$, there is a point $k\geq
h$, such that at every point $p\geq k$ in $\mathcal{R}$, for every
$A_i$ such that $a\in HB_i$, $a\in M_{i,p}$ iff $a\in
\sembr{P_\mathcal{A}}$.
\end{lemma}

\begin{proof*}

We prove by induction on $\pi(a)$. For each i, let $HBI_i =
head(IDB_i)$.

\begin{itemize}
\item
\textit{Base case:} $\pi(a)=0$ ($a$ is a leaf in the dependency
graph of $P_\mathcal{A}$).

Let $A_i$ be an agent with $a\in HB_i$. There are three cases:
    \begin{enumerate}

    \item $a\in HBI_i$. There must be a clause of the
    form $a\leftarrow$ in $IDB_i$. $a\leftarrow$ is also in $IDB_\mathcal{A}$.
    At every point $m\geq 0$, $a\in M_{i,m}$ and $a\in \sembr{P_\mathcal{A}}$.

    \item $a\in HBE_i$.
    There is no change in the environment after $h$, at every point $k\geq h$,
    $a\in M_{i,k}$ iff $a\in EDB_{i,k}$ iff $a\in \sembr{P_\mathcal{A}}$.

    \item $a\in HIN_i$. There must be an agent $A_j$ such that
    $D(i,j)\neq\emptyset$ and $a\in HBE_j\cup HBI_j$.
    By definition \ref{def:run} of the run,
    there must be a point $p\geq h$ such that there is a transition
    \[
    \triangle_{p}\xrightarrow{j\leadsto i}\triangle_{p+1}
    \]
    Moreover, every transition that can delete (or insert) $a$
    from (or into) $IN_i$ after point $h$ must also
    have the form $\triangle_{q}\xrightarrow{j\leadsto
    i}\triangle_{q+1}$ for some $A_j$ such that $D(i,j)\neq \emptyset$
    and $a\in HBE_j\cup HBI_j$.
    By the definition of transition of the form
    $\triangle\xrightarrow{j\leadsto i}\triangle'$
    in definition \ref{def:run} and the operator
    $Upa$ in section \ref{subsec:CommunicationSensing}, for a
    transition $\triangle_{p}\xrightarrow{j\leadsto
    i}\triangle_{p+1}$, $A_i$ will update $IN_i$ as follows
        \[
        IN_{i,p+1} = (IN_{i,p} \setminus D(i,j))\cup S
        \]
    where $S=D(i,j)\cap M_{j,p}$. Since $a\in D(i,j)$, $a\in M_{i,p+1}$ iff $a\in IN_{i,p+1}$ iff $a\in
    M_{j,p}$. As shown in $1$ and $2$,
    at every point $k\geq h$, for every $A_j$ such that $a\in HBI_j\cup HBE_j$,
    $a\in M_{j,k}$ iff $a\in \sembr{P_\mathcal{A}}$. So at every point $k\geq p$, $a\in
    M_{i,k+1}$ iff $a\in \sembr{P_\mathcal{A}}$.
    \end{enumerate}

We have proved that for each $A_i$ such that $a\in HB_i$ there a
point $p_i$ such that at every point $k\geq p_i$, $a\in M_{i,k}$
iff $a\in \sembr{P_\mathcal{A}}$. Take $p=max(p_1,\dots, p_n)$. At
every point $k\geq p$, for every agent $A_i$ such that $a\in
HB_i$, $a\in M_{i,k}$ iff $a\in \sembr{P_\mathcal{A}}$.

\item
\textit{Inductive case:} Suppose the lemma holds for every atom $a$
with $\pi(a)\leq m$, $m\geq 0$. We show that the lemma also holds
for $a$ with $\pi(a)=m+1$.

Let $A_i$ be an agent with $a\in HB_i$. Clearly $a\not\in
HBE\supseteq HBE_i$. There are two cases:

\begin{enumerate}

\item
$a\in HBI_i$. The atom dependency graph of $P_\mathcal{A}$
is acyclic, every child $b$ of $a$ has $\pi(b)\leq m$. By the
inductive assumption, for each $b$ there is a point $p_b$ such
that at every point $k\geq p_b$, $b\in M_{i, p_b}$ iff $b\in
\sembr{P_\mathcal{A}}$. The set of children of $a$ in the atom
dependency graph of $P_\mathcal{A}$ is the same as the set of
atoms in the body of all clauses of the definition of $a$.  As
$IDB_\mathcal{A}$ is bounded, $a$ has a finite number of children
in the atom dependency graph of $P_\mathcal{A}$ and the definition
of $a$ is finite. Let $p_a$ is the maximum number in the set of
all such above $p_b$ where $b$ is a child of $a$. At every point
$k\geq p_a$, for every child $b$ of $a$, by the inductive
assumption, $b\in M_{i,k}$ iff $b\in \sembr{P_\mathcal{A}}$. We
prove that $a\in M_{i,k}$ iff $a\in\sembr{P_\mathcal{A}}$.

By lemma \ref{lemma:app10}, $a\in M_{i,k}$ iff there is a rule
$a\leftarrow Bd$ in $P_{i,k}=IDB_i\cup EDB_{i,k}\cup IN_{i,k}$
such that $M_{i,k}\models Bd$. By inductive assumption for every
$b\in atom(Bd)$, $b\in M_{i,k}$ iff $b\in \sembr{P_\mathcal{A}}$.
Moreover $a\leftarrow Bd$ is also a rule in $P_\mathcal{A}$. Thus
$a\in M_{i,k}$ iff there is a rule $a\leftarrow Bd$ in
$P_\mathcal{A}$ such that $\sembr{P_\mathcal{A}}\models Bd$ iff
$a\in \sembr{P_\mathcal{A}}$ (by lemma \ref{lemma:app10}).

\item
$a\in HIN_i$. As shown in 1, for every $A_j$ such that $a\in HBI_j$
there is a point $p_j$, such that at every point $k\geq p_j$,
$a\in M_{j,k}$ iff $a\in\sembr{P_\mathcal{A}}$. Let $p$ be the
maximum of all such $p_j$. Clearly, at every point $k\geq p$, for
every $A_j$ such that $a\in HBI_j$, $a\in M_{j,k}$ iff
$a\in\sembr{P_\mathcal{A}}$.

Follow similarly as case $3$ in base case of the proof, there is a
point $p'\geq p+1$ such that at every point $k\geq p'$, $a\in
M_{i,k}$ iff $a\in M_{j,k}$. It also means that at every point
$k\geq p'$, $a\in M_{i,k}$ iff $a\in\sembr{P_\mathcal{A}}$.
\end{enumerate}

We have proved that for each $A_i$ such that $a\in HB_i$ there a
point $p_i$ such that at every point $k\geq p_i$, $a\in M_{i,k}$
iff $a\in \sembr{P_\mathcal{A}}$. Take $p=max(p_1,\dots, p_n)$. At
every point $k\geq p$, for every agent $A_i$ such that $a\in
HB_i$, $a\in M_{i,k}$ iff $a\in \sembr{P_\mathcal{A}}$.
\end{itemize}
\begin{flushright} $\square$ \end{flushright}
\end{proof*}

\section{Proof of theorem \ref{theo:2}} \label{app:ProofTheorem2}

Let $\mathcal{A}$ be an IO-acyclic and IO-finite multiagent
system. Obviously $\mathcal{A}$ is also bounded. Let $\mathcal{R}$
be a run of $\mathcal{A}$. By theorem \ref{theo:1}, $\mathcal{R}$
is convergent. By lemma \ref{lemma:app12}, for every atom $a$ in
$G_\mathcal{A}$ there is a point $k_a$ such that at every point
$p\geq k_a$, for every agent $A_i$ such that $a\in HB_i$, $a\in
M_{i,p}$ iff $a\in \sembr{P_{\mathcal{A}}}$. As $G_\mathcal{A}$ is
finite, take the largest number $k$ of all such $k_a$'s for every
atoms $a$ in $G_\mathcal{A}$. Obviously, at every point $p\geq k$,
for every agent $A_i$, $M_{i,k}=M_{i,p}$. Thus $\mathcal{R}$ is
strongly convergent. The system is stabilizing and theorem
\ref{theo:2} follows immediately.

\bibliographystyle{acmtrans}

\end{document}